\documentclass[onecolumn,showpacs,preprintnumbers,amsmath,amssymb, longbibliography]{revtex4}
\usepackage{graphicx}
\usepackage{dcolumn}
\usepackage{bm}
\usepackage{epsfig}
\usepackage{subfigure}

\newcommand{\bi}{\begin{itemize}}
\newcommand{\ei}{\end{itemize}}
\newcommand{\be}{\begin{eqnarray}}
\newcommand{\ee}{\end{eqnarray}}
\newcommand{\beq}{\begin{equation}}
\newcommand{\eeq}{\end{equation}}
\newcommand{\dd}{\text{d}}

\newcommand{\bbmatrix}{\left( \begin{array}}
\newcommand{\eematrix}{\end{array} \right)}

\begin{document}

\title{Application of Pontryagin's Minimum Principle to Grover's Quantum Search Problem}

\author{Chungwei Lin$^1$\footnote{clin@merl.com}, Yebin Wang$^1$, Grigory Kolesov$^{1,2}$, Uro\v{s} Kalabi\'c $^1$}
\date{\today}
\affiliation{$^1$Mitsubishi Electric Research Laboratories, 201 Broadway, Cambridge, MA 02139, USA \\
$^2$Harvard University, Cambridge, MA 02138, USA}

\begin{abstract}
Grover's algorithm is one of the most famous algorithms which explicitly demonstrates how the quantum nature can be utilized to accelerate the searching process. In this work, Grover's quantum search problem is mapped to a time-optimal control problem. Resorting to Pontryagin's Minimum Principle we find that the time-optimal solution has the bang-singular-bang structure. This structure can be derived naturally, without integrating the differential equations, using the geometric control technique where Hamiltonians in the Schr\"odinger's equation are represented as vector fields. In view of optimal control, Grover's algorithm uses the bang-bang protocol to approximate the optimal protocol with a minimized number of bang-to-bang switchings to reduce the query complexity. Our work provides a concrete example how Pontryagin's Minimum Principle is connected to quantum computation, and offers insight into how a quantum algorithm can be designed. 
\end{abstract}

%\pacs{31.15.A-,71.55.-i,73.20.hb}
\maketitle

%%%%%%%%%%%%%%%%%%%%%%%%%%%%%%%%%%%%%%%%%%%%%%%%%%%%%%%%%%%%%%%%%%%%%%%%%%%%
\section{Introduction} 

Quantum computation deliberately uses the quantum-mechanical phenomena, such as superposition and entanglement, to reduce the computation time or the number of queries to accomplish certain tasks \cite{Neilson_book, book:Kaye_book}. Well known quantum algorithms include Shor's algorithm for factoring \cite{Shor:1997:PAP:264393.264406} and Grover's algorithm for searching an unstructured database or an unordered list \cite{Grover:1996:FQM:237814.237866, PhysRevLett.79.325}. 
%%%%%%%%
While the standard paradigm for quantum computation involves a {\em discrete} sequence of unitary logic gates \cite{Deutsch_Jozsa_92, Simon:1997:PQC:264393.264405, Shor:1997:PAP:264393.264406, Grover:1996:FQM:237814.237866, PhysRevLett.79.325, Bennett_97, PhysRevLett.103.150502} there exists another paradigm, pioneered by Farhi and Gutmann \cite{PhysRevA.57.2403}, where the quantum register evolves under some designed Hamiltonian which can vary {\em continuously} in time \cite{PhysRevA.57.2403, PhysRevE.58.5355, Farhi472, PhysRevA.65.042308}. The concept of 
``continuous-time'' quantum computation explicitly allows established physics principles such as the adiabatic theorem \cite{Farhi_00, PhysRevLett.103.080502, PhysRevA.90.052317} and the Trotter-Suzuki decomposition \cite{Trotter-1959, Suzuki1976} to guide how quantum algorithms can be designed. The adiabatic theorem is, for example, the foundation of the quantum annealing technique \cite{PhysRevE.58.5355, Brooke-1999, Santoro-2006, RevModPhys.80.1061, Johnson_2010} and the fast, non-adiabatic evolution is found to be helpful for other problems \cite{PhysRevLett.115.230501, Heim215}. More recently, there are quantum algorithms based on the variational principle, notably the Variational Quantum Eigensolver (VQE) \cite{Peruzzo-2014, PhysRevX.6.031007, Kandala-2017} and the Quantum Approximate Optimization Algorithm (QAOA) \cite{Farhi_14, PhysRevA.92.042303, Lin_Lin_16}. They are more fault-tolerant than quantum algorithms of the standard paradigm and are promising for Noisy Intermediate-Scale Quantum (NISQ) technology \cite{Preskill2018quantumcomputingin}. %The variational based algorithms are hybrid in nature: they use quantum computer to compute a pre-defined cost function and the classical algorithm for optimization. 

Generally, when applying quantum algorithms to solve a classical NP (non-deterministic polynomial-time) problem, we are given a quantum ``problem Hamiltonian'' (oracle) whose ground state is the solution of the original classical problem \cite{RevModPhys.80.1061, PhysRevLett.108.130501, Crosson_14, PhysRevA.94.022309}. Designing a quantum algorithm is equivalent to find a ``driving Hamiltonian'' and an initial state, both easily implemented, that can steer the initial state to the target state (e.g., ground state of the problem Hamiltonian) within the shortest time. From this point of view, time-optimal control \cite{book:Luenberger, book:Liberzon, book:GeometricOptimalControl} appears to be fundamentally connected to quantum computation as both address (i) if the target state can be reached and (ii) how to reach the target state in the shortest time. 
Recently, Pontryagin's Minimum Principle (PMP) \cite{book:Pontryagin} has been applied to quantum state preparation \cite{PhysRevA.97.062343} and non-adiabatic quantum computation \cite{PhysRevX.7.021027}. Due to the linearity of Schr\"odinger's equation, time-optimal control generally has the bang-bang form, i.e., the control takes its extreme values. Indeed, Ref.~\cite{PhysRevA.97.062343} shows that the bang-bang protocol takes the minimum time to drive a two-qubit product state to the fully entangled state; Ref.~\cite{PhysRevX.7.021027} establishes the connection between the bang-bang protocol and the QAOA algorithm by solving spin-glass problems; Ref.~\cite{PhysRevA.98.012301} demonstrates that pulses of bang-bang type take the minimum time to cancel the second-order noise. 
In this paper, we formulate Grover's problem as a time-optimal control problem and apply PMP to solve it. By recasting the problem involving only observable dynamical variables, we are able to show that the time-optimal control is of bang-singular-bang type without explicitly 
solving the necessary optimality conditions derived from PMP. 
%integrating a differential equation. 
In the language of control theory, Grover's algorithm can be regarded as a protocol that approximates the optimal control by the bang-bang control with a minimum number of switchings. Our analysis directly applies to the system effectively involving a single qubit \cite{PhysRevA.97.062343, PhysRevLett.114.100801, PhysRevLett.118.010501}, and may provide insight into the problems of higher dimensions \cite{PhysRevX.8.031086, PhysRevLett.122.020601}.

The rest of the paper is organized as follows. In Section II we formulate the problem and state the necessary conditions for a time-optimal solution derived from PMP. In Section III we show that the bang-singular-bang structure satisfies the necessary conditions and compares this solution to other protocols including the pure singular control %\cite{PhysRevA.57.2403}
and the pure bang-bang control which turns out to be Grover's algorithm. %\cite{PhysRevLett.79.325}. 
Advantages of numerically applying PMP to problems of higher dimensions are discussed. In Section IV we re-formulate the problem in terms of geometric control, and show that the time-optimal bang-singular-bang control can be naturally derived in this formalism. The relation to quantum state preparation are pointed out. Section V is the conclusion. In the Appendix, a few detailed steps regarding the geometric control technique are provided. 

%%%%%%%%%%%%%
\section{Problem statement and Pontryagin's Minimum Principle}

In this section we formulate Grover's problem in terms of two Hamiltonians and summarize the relevant results from PMP. Because both quantum mechanics and PMP use the term ``Hamiltonian'', we shall use ``Hamiltonian'' (symbol $H$) in the quantum-mechanical sense, i.e., it is a matrix that governs the dynamics of the wave function; use ``c-Hamiltonian'' (symbol $\mathcal{H}$) to represent the control-Hamiltonian, a scalar function defined in control theory.

\subsection{Problem statement}
Following Ref. \cite{PhysRevA.57.2403, PhysRevA.65.042308}, Grover's problem can be formulated using Hamiltonians. 
Given the ``problem Hamiltonian'' $H_w = |w \rangle \langle w|$ and the ``driving Hamiltonian'' $H_s = |s \rangle \langle s|$ where $\langle s| w \rangle \equiv x < 1$ %(no summation over $w$ or $s$)
($x$ and $|s\rangle$ are assumed to be real for now, and $|s\rangle$ for Grover's problem is given in Eq.~\eqref{eqn:initial}), we want to find the time-optimal protocol $u(t)$ that brings the initial state $| \psi_i \rangle = |s \rangle$ to the target state $|\psi_{target} \rangle = |w \rangle$ for the system evolved under a time-dependent Hamiltonian 
\beq 
\begin{aligned}
H(t; u) & = \frac{1}{2} ( H_w + H_s)  + u(t) \,\frac{1}{2} (H_w- H_s) \\
&\equiv H_0 + u(t) H_d,  \text{ with } |u(t)| \leq 1.
\end{aligned}
\label{eqn:Problem}
\eeq 
%In Eq.~\eqref{eqn:Problem}, $u$ is a scalar we are able to control. Note that 
The control $u(t)$ of the problem is assumed to be bounded by $|u| \leq 1$ \cite{convention}. 
In this paper, the terms ``control'', ``protocol'', and ``algorithm'' are  used interchangeably; a given control/protocol/algorithm corresponds to a specific $u(t)$.  Because all non-target states are {\em degenerate} in energy, we can express both Hamiltonians using two orthogonal states $|w \rangle$, $|\bar{w} \rangle = \left[|s\rangle - x | w \rangle \right]/\sqrt{1-x^2}$. The initial state is given by $| \psi_i \rangle = |s \rangle = x | w \rangle + \sqrt{1-x^2} | \bar{w} \rangle$. In terms of Pauli matrices defined as 
\beq 
\sigma_x = \begin{bmatrix} 0&1\\1&0 \end{bmatrix},\,\,
\sigma_y = \begin{bmatrix} 0&-i\\i&0 \end{bmatrix},\,\,
\sigma_z = \begin{bmatrix} 1&0\\0&-1 \end{bmatrix},  
\nonumber %\label{eqn:Pauli}
\eeq 
the Hamiltonians in Eq.~\eqref{eqn:Problem} are
\begin{subequations}
\begin{align}
H_w &=  \frac{1}{2} (e + \sigma_z), \\ 
H_s &= \frac{1}{2} e + \frac{1}{2} \left[ (2x \sqrt{1-x^2}) \sigma_x + (2x^2-1) \sigma_z \right]  
\label{eqn:Hs_grover}\\ 
%%%
H_0 &  =  \frac{1}{2} e + \frac{x}{2} \left[ \sqrt{1-x^2}  \sigma_x + x \sigma_z \right] ,
\\
%%%
H_d &= -\frac{x}{2} \sqrt{1-x^2}  \sigma_x + \frac{1}{2} (1-x^2) \sigma_z ,
\\
%%%%
| \psi_i \rangle &= \begin{bmatrix} x \\ \sqrt{1-x^2} \end{bmatrix}, 
\,\,\, 
| \psi_{target} \rangle = \begin{bmatrix} 1 \\ 0\end{bmatrix}. 
\label{eqn:initial_target}
\end{align}
\label{eqn:H_grover}
\end{subequations}
Note that the target state is allowed to have an arbitrary phase.
For later discussions, we further define 
\beq 
\begin{aligned}
H_X &= H_0 - H_d = H_s, \\
H_Y &= H_0 + H_d = H_w,
\end{aligned}
\label{eqn:H_XY}
\eeq 
i.e., $H_X$ corresponds to $u=-1$ whereas $H_Y$ to $u=+1$. The initial state is chosen to be 
\beq 
| s \rangle = \frac{1}{\sqrt{N}} \sum_{k=1}^N | k \rangle,
\label{eqn:initial}
\eeq 
with $N$ the dimension of search space (Hilbert space) spanned by $\{ |k \rangle \}$. As  Hamiltonians given in Eqs.~\eqref{eqn:H_grover} are just $2 \times 2$ matrices, the  problem dimension $N$ is encoded in the overlap $x = 1/\sqrt{N}$.  Throughout this paper, $x$ (different from $\pmb{x}$) exclusively represents the overlap between the initial and the target state $\langle s| w \rangle$. The two-dimensional wave function will be referred to as a ``qubit'' state.

\subsection{Time-optimal control problem and Pontryagin's Minimum Principle}

The necessary conditions for a time-optimal solution derived from PMP are discussed in this subsection. 
We focus on the control-affine control system, where the dynamics of its ``state variables'' $\pmb{x}$ are described by 
\beq 
 \dot{ \pmb{x} } = \mathbf{f}(\pmb{x}) + u(t) \,\mathbf{g}(\pmb{x}),\,\, \pmb{x} \in R^n, u \in R
 \label{eqn:control-affine}
\eeq  
To differentiate from ``quantum state'', $\pmb{x}$ will be referred to as ``dynamical variables''; $\mathbf{f}$ and $\mathbf{g}$ are smooth vector fields which are functions of $\pmb{x}$. Control-affine system is directly related to the Schr\"odinger's equation. In the language of differential geometry, $\pmb{x}$ defines a manifold; $\mathbf{f}$ and $\mathbf{g}$ belong to the tangent space of that manifold. We assume the admissible range of $u$ is bounded by $|u| \leq 1$. Time-optimal control problem is defined as follows: given Eq.~\eqref{eqn:control-affine}, find the optimal control $u(t)$ to minimize the cost function  
\beq 
J = \lambda_0 \int_0^{t_f} \dd t + \mathcal{C}(\pmb{x}(t_f) ),
\label{eqn:J}
\eeq  
where $t_f$ is the final time, $\lambda_0$ is a positive constant, and $\mathcal{C}(\pmb{x}(t_f))$ is a terminal cost function depending only on the values of the dynamical variables at $t_f$. To make sense of a ``time-optimal'' solution, $\lambda_0$ has to be positive. We only consider the time-invariant problem where $\mathbf{f}$, $\mathbf{g}$, and $\mathcal{C}$ do not depend explicitly on time $t$.

Following PMP \cite{book:Luenberger}, a control-Hamiltonian is defined as 
\beq 
\begin{aligned}
\bar{ \mathcal{H} }_c (t)
&= \lambda_0 + \langle \pmb{\lambda}(t), \mathbf{f}(\pmb{x}) \rangle + u(t) \langle \pmb{\lambda}(t), \mathbf{g}  (\pmb{x}) \rangle \\
&\equiv \lambda_0 + \langle \pmb{\lambda}(t), \mathbf{f}(\pmb{x}) \rangle + u(t) \Phi(t) \\
&\equiv \lambda_0 + \mathcal{H}_c (t).
\end{aligned}
\label{eqn:c-Hamil_01}
\eeq 
$\pmb{\lambda}$ is referred to as a set of ``costate'' variables (or the conjugate momentum), which has the same dimension of $\pmb{x}$. $\langle \cdot, \cdot \rangle$ is the inner product introduced for two real-valued vectors. A switching function $\Phi(t) $ is defined as 
\beq 
\Phi(t) = \langle \pmb{\lambda}(t), \mathbf{g}  (\pmb{x}) \rangle, 
\label{eqn:switch_func_01}
\eeq 
which plays the most important role in determining the structure of optimal control. 

Given an optimal solution $(\pmb{x}^*, \pmb{\lambda}^*; u^*)$ to the time-optimal control problem, it has to satisfy the following necessary conditions:
\begin{subequations}
\begin{align} 
& \dot{ \pmb{x} }^*(t) = + \left( \nabla_{\pmb{\lambda}} \mathcal{H}_c\right), \,\,\,\pmb{x}^*(0) \text{ is given.}  
\label{eqn:original_dynamics}
\\
%%%
& \dot{ \pmb{\lambda} }^*(t) = - \left( \nabla_{\pmb{x}} \mathcal{H}_c\right)^T, \,\,\, \pmb{\lambda}^*(t_f) =  \nabla_{\pmb{x}} \mathcal{C} |_{ \pmb{x}^*(t_f) }\label{eqn:costate_dynamics} \\
%%% & \pmb{\lambda}(t_f) =  \nabla_{\pmb{x}} \Psi |_{t=t_f},  \label{eqn:costate_tf} \\
%%% 
& \bar{ \mathcal{H} }_c = 
\lambda_0 + \mathcal{H}_c = \text{ const. } 
\label{eqn:const_c-H} \\ 
& u^*(t) = \begin{cases} +1 & \text{ if } \Phi(t)<0 \\
          -1 & \text{ if } \Phi(t)>0      \\
          \text{undetermined} & \text{ if } \Phi(t)=0 \end{cases}.
\label{eqn:bang_condition}
\end{align}
\label{eqn:necc_cond_state_variables}
\end{subequations}
Let us elaborate on these necessary conditions. Eq.~\eqref{eqn:original_dynamics} is identical to the dynamics defined in Eq.~\eqref{eqn:control-affine}. Eq.~\eqref{eqn:costate_dynamics} defines the dynamics of costate variables, whose boundary condition is fixed at the final time $t_f$. Eq.~\eqref{eqn:const_c-H} holds for the time-invariant problem. If the final time $t_f$ is not fixed (i.e., $t_f$ is allowed to vary to minimize $\mathcal{C}$), then $\bar{ \mathcal{H} }_c = 0 $. Because $\lambda_0$ is a positive constant, we conclude that $\mathcal{H}_c(t)$ has to be a negative constant for a time-optimal solution. In the following analysis, we shall focus on $\mathcal{H}_c$ (instead of $\bar{ \mathcal{H} }_c$), and the condition \eqref{eqn:const_c-H} is replaced by 
\beq 
\mathcal{H}_c = \text{const.} \leq 0 .
\label{eqn:negative_c-H}
\eeq 
Eq.~\eqref{eqn:bang_condition} means that the optimal control takes the extreme values ($\pm 1$ in this case) when the switching function is nonzero. This structure is referred to as ``bang-bang'' protocol. If $\Phi(t) = 0$ over a finite interval of time, then $u$ is undetermined from Eq.~\eqref{eqn:bang_condition} and more analysis is needed. When $\Phi=0$, the time-optimal control $u^*$ may not take its extreme values and is referred to as a singular control. The behavior of $\Phi=0$ will be analyzed in Section \ref{Sec:singular} to establish the structure of time-optimal solution.

\subsection{Application to the Schr\"odinger's equation}

In Grover's problem, the dynamics of the wave function obey the Schr\"odinger's equation
\beq 
\begin{aligned}
i \frac{\dd}{\dd t} | \Psi(t) \rangle &= 
\left[ H_0 + u(t) \, H_d \right] | \Psi(t) \rangle, \\ \text{with }
| \Psi(t) \rangle  
&= \begin{bmatrix} \Psi_0 \\ \Psi_1 \end{bmatrix}
= \begin{bmatrix} \Psi_{0,R} + i \Psi_{0,I} \\ 
\Psi_{1,R} + i \Psi_{1,I} \end{bmatrix}.
\end{aligned}
\label{eqn:dynamics_grover_explicit}
\eeq 
The initial and final states are given in Eq.~\eqref{eqn:initial_target}. To make the final state as close to $[1,0]^T$ as possible, the terminal cost function can be chosen as 
\beq 
\mathcal{C}( \Psi(t_f) ) = -\frac{1}{2} |\Psi_0 (t_f) |^2.
\label{eqn:cost_Grover}
\eeq 
As explicitly shown in Eq.~\eqref{eqn:dynamics_grover_explicit}, $|\Psi \rangle$ contains four real variables. As a result costate variables $\Pi_{0,R}$,  $\Pi_{0,I}$, $\Pi_{1,R}$,  $\Pi_{1,I}$ are needed. It is convenient to express the four costate variables as a two-dimensional complex vector
\beq 
| \Pi(t) \rangle = \begin{bmatrix} \Pi_{0,R} + i \Pi_{0,I} \\ 
\Pi_{1,R} + i \Pi_{1,I} \end{bmatrix}.
\eeq 
Using the property that $H_0$ and $H_d$ are real-valued, we can express the c-Hamiltonian and switching function as 
\beq 
\begin{aligned}
\mathcal{H}_{Q,c} &= \text{Im} \langle \Pi(t) | 
\left[ H_0 + u(t) \, H_d \right] | \Psi(t) \rangle, \\
\Phi_Q(t) &=\text{Im} \langle \Pi(t) |  H_d  | \Psi(t) \rangle.
\end{aligned} 
\label{eqn:quantum_Hoc_switch}
\eeq 
Applying Eq.~\eqref{eqn:costate_dynamics}, one can derive that the dynamics of $|\Pi(t) \rangle$ are governed by the same Schr\"odinger's equation, with the boundary condition given at $t_f$ \cite{PhysRevA.97.062343, PhysRevX.7.021027}: 
\beq 
\begin{aligned}
i \frac{\dd}{\dd t} | \Pi(t) \rangle = 
\left[ H_0 + u(t) \, H_d \right] | \Pi(t) \rangle, \\
\text{ with }|\Pi(t_f) \rangle = - \begin{bmatrix} \Psi_0(t_f) \\ 0 \end{bmatrix}.
\end{aligned}
\label{eqn:Pi_dynamics}
\eeq 
The condition ~\eqref{eqn:bang_condition} still holds, with the switching function computed using Eq.~\eqref{eqn:quantum_Hoc_switch}. The subscript `Q' in Eqs.~\eqref{eqn:quantum_Hoc_switch} indicates `quantum' and will be dropped from now on. 

Eqs.~\eqref{eqn:dynamics_grover_explicit} and ~\eqref{eqn:cost_Grover} allow us to do the bruteforce optimization to determine the optimal $u(t)$. Eqs.~\eqref{eqn:quantum_Hoc_switch} and ~\eqref{eqn:Pi_dynamics} allow us to check if a control $u(t)$ satisfies the necessary conditions. One may expect that the optimal control is mostly of bang-bang type, as $\Phi(t)=0$ can only occupy a negligible region in the manifold. In Section \ref{sec:optimal_protocol} we numerically show that the optimal protocol actually has the bang-singular-bang structure. 

%%%%%%%%%%%%%
%\section{Grover's problem}

%%%%%%%%%%%%%%%%%%%
\section{Optimal protocol \label{sec:optimal_protocol}}
In this section we show that the time-optimal control has the bang-singular-bang structure. In particular, we numerically check that a solution bearing this structure satisfies all necessary conditions imposed by PMP, and explicitly show that the time-optimal solution does take a shorter time to reach the target state when compared to other known protocols.

\subsection{Two existing protocols \label{sec:existing}}

We begin the discussion by introducing two existing protocols which will be compared to the optimal protocol described shortly. The first one is provided by Farhi and Gutmann ~\cite{PhysRevA.57.2403}. By choosing $u(t) = 0$ in Eq.~\eqref{eqn:Problem}, the wave function is 
\beq 
\begin{aligned}
| \Psi (t) \rangle &= e^{-i \frac{1}{2} (H_s + H_w) t} |s \rangle \\
&= e^{-i \frac{t}{2} } 
\begin{bmatrix} x \cos \frac{xt}{2} - i \sin \frac{xt}{2} \\ 
\sqrt{1-x^2} \cos \frac{xt}{2} \end{bmatrix}
\equiv \begin{bmatrix} \Psi_0(t) \\ \Psi_1(t) \end{bmatrix} .
\end{aligned}
\eeq 
As $ | \Psi_0(t) |^2 = x^2 \cos^2 \frac{xt}{2} + \sin^2 \frac{xt}{2}$, we find that for total computation time $t_f = \frac{\pi}{x}$, $ | \Psi_0 (t_f)|^2 = 1$. This protocol will be referred as the ``singular'' control as $u$ does not take either of its extreme values. 
The second protocol is the celebrated Grover's algorithm \cite{PhysRevLett.79.325}, where the initial state $| s \rangle$ is evolved by alternating $U_w = 2 | w \rangle \langle w| - I $ and $U_s = I - 2 | s\rangle \langle s|$ about $N = \frac{\pi}{4x} $ times, i.e.,
\beq 
| \psi_f \rangle = (U_s U_w)^N | s \rangle.
\eeq 
In terms of Eq.~\eqref{eqn:Problem}, $U_w$ is achieved by $e^{-i \pi H_w}$ and $U_s$ by $e^{-i \pi H_s}$ up to an irrelevant sign. The total computation time for Grover's algorithm is therefore $(\pi+ \pi) N \sim \frac{\pi^2}{2x}$. Grover's algorithm and the singular protocol have the same asymptotic $x^{-1}$ behavior, but the latter has a smaller prefactor. The annealing algorithm proposed by Roland and Cerf \cite{PhysRevA.65.042308} is slower but more robust, as it applies the adiabatic theorem concerning the (instant) gap between the  ground and the first excited state, but does not explicitly use the structure that all non-target states are degenerate. The annealing algorithm will not be discussed here. However, we note that the optimal $u(t)$ obtained using PMP is non-adiabatic, and the procedure does not know if there is a gap at any instant of time, which is essential for the adiabatic quantum computing.

\subsection{Optimal control and analytical analysis} 

\begin{figure}[ht]
\begin{center}
\includegraphics[width=0.5\textwidth]{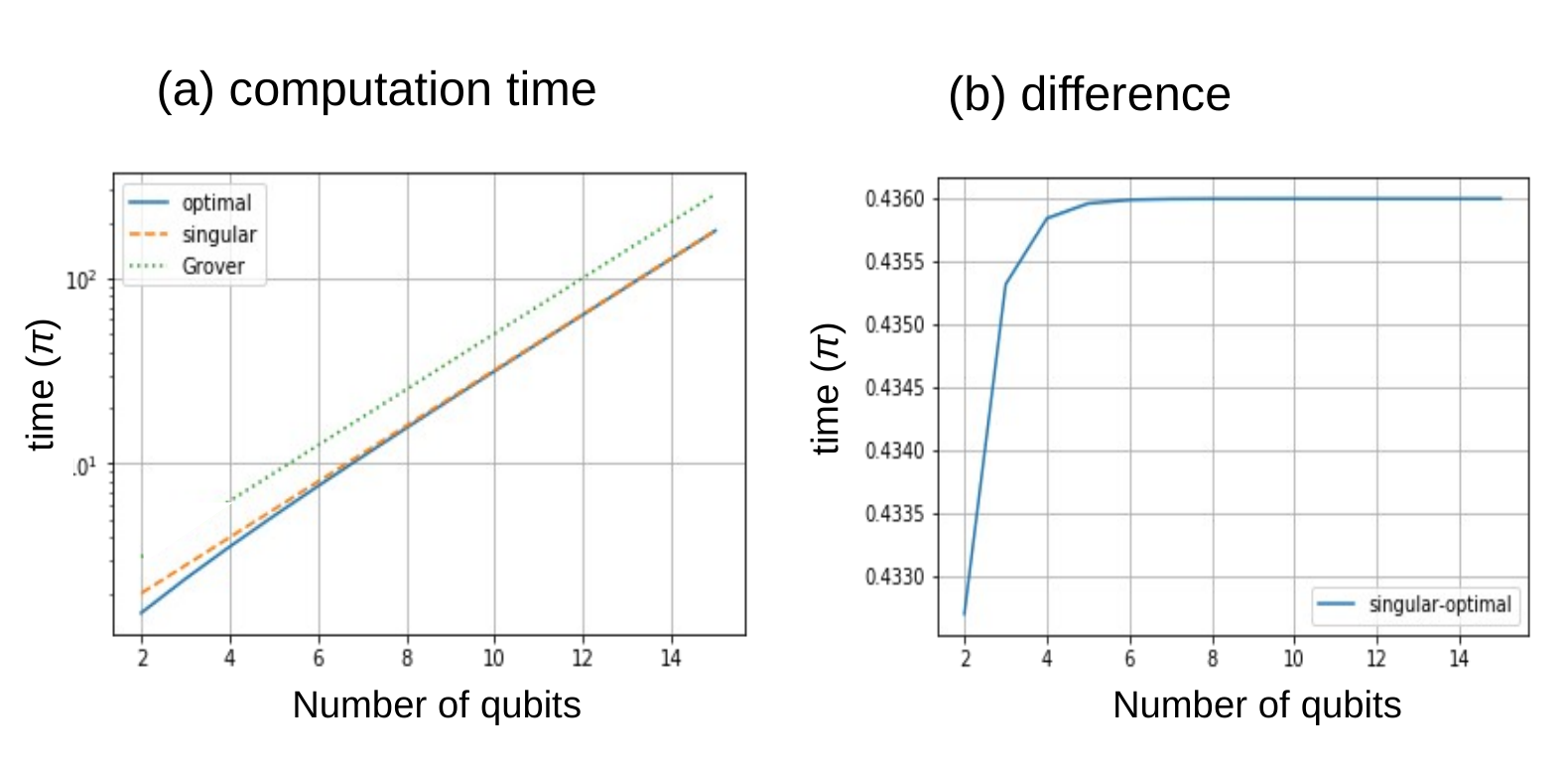}
\caption{(Color Online)  (a) The total computation time for the optimal control, the singular control, and Grover's algorithm in logarithmic scale. For $n$ number of qubits, the overlap $x = 2^{-n/2}$. For the smaller system, the optimal control has a noticeable speedup. When the system becomes large, these two methods are almost equivalent. Grover's algorithm has the same asymptotic behavior with a larger prefactor. (b) the difference of computation time  between the optimal and the singular control in linear scale. The optimal control takes less time; the difference approaches 0.436 $\pi$ at large $n$.
}
\label{fig:time_optimal}
 \end{center}
\end{figure}

We claim that the time-optimal control has the bang-singular-bang structure which will be proved using geometric control in Section \ref{sec:geometric_control}. In particular, the order is $u=+1$ (bang), $u=0$ (singular), and $u=-1$ (bang). The time periods for the initial and the final bang control are identical because the problem is unchanged if we swap the initial and target states. %This conclusion was first conjectured from extensive numerical experiments. 
Let us first investigate the consequences of applying this protocol. 

Assume that the time intervals of two bang controls are both $t_1$, and that of the singular control is $t_2$. The wave function at $t_f = 2t_1+t_2$ is given by 
\beq 
\begin{aligned}
|\Psi(t_1; t_2) \rangle
&=   e^{-i t_1 H_s} e^{-i \frac{t_2}{2}(H_s+H_w)} e^{-i t_1 H_w} |s \rangle
\equiv \begin{bmatrix}  \Psi_0 (t_1, t_2 ) \\ \Psi_1 (t_1, t_2 )
\end{bmatrix}
\end{aligned}
\label{eqn:psi_bsb_grover}
\eeq 
A straightforward calculation gives (neglecting the phase factor)
\beq 
\begin{aligned}
\Psi_1 (t_1, t_2 ) %&= \sqrt{1-x^2} \left(
%\cos^2 \frac{t_1}{2} \cos \frac{x t_2}{2} + (1-4 x^2) \cos \frac{x t_2}{2} \sin^2 \frac{t_1}{2} - 4 x \cos \frac{t_1}{2} \sin \frac{t_1}{2} \sin \frac{x t_2}{2} \right) \\
%%
&= \sqrt{1-x^2} \left[ \cos \frac{x t_2}{2} (1-4x^2 \sin^2 \frac{t_1}{2} )-  2x \sin \frac{x t_2}{2} \sin t_1 \right].
\end{aligned}
\eeq 
Note that maximizing $|\Psi_0 (t_1, t_2 )|^2$ is equivalent to minimizing $|\Psi_1 (t_1, t_2 )|^2$. If the target state can be exactly reached,  $\Psi_1 (t_1, t_2 )=0$ which allows us to represent $t_2$ as a function of $t_1$: 
\beq 
\tan \frac{x t_2}{2} = 
\frac{1-4 x^2 \sin^2 \frac{t_1}{2} }{ 2x \sin t_1 }
= \frac{1-2 x^2  + 2x^2 \cos t_1 }{ 2x \sin t_1 }.
\label{eqn:t2(t1)}
\eeq 
%This equation gives $t_2 (t_1)$ and we restrict $t_1 \geq 0$.
As a result the total computation time can be written as a function of $t_1$:
\beq 
\begin{aligned}
t_f (t_1) = 2t_1 + t_2(t_1).
\end{aligned} 
\eeq 
A time-optimal solution requires $\frac{\dd t_f}{\dd t_1} = 2 + \frac{\dd t_2}{\dd t_1}    = 0 $, from which we get $\frac{\dd t_2}{\dd t_1} = -2$.
%\beq  
%= 2 + \frac{dt_2}{dt_1}  
%\Rightarrow .
%\label{eqn:optimal_0}
%\eeq 
To get the optimal $t_1$, we take $\frac{\dd}{\dd t_1}$ on both sides of Eq.~\eqref{eqn:t2(t1)}:
\beq 
\begin{aligned}
&\sec^2 (\frac{x t_2}{2}) \cdot \frac{x}{2} \cdot \frac{\dd t_2}{\dd t_1}
=\frac{1}{\cos^2 (\frac{x t_2}{2})} \cdot \frac{x}{2} \cdot \frac{\dd t_2}{\dd t_1} \\
%= \frac{ -x\left[ 4x^2 + 2 \cos t_1 (1-2x^2) \right]  }{ 4x^2 \sin^2 (t_1) } \\
%%% 
%\Rightarrow &  \frac{ (1-2 x^2  + 2x^2 \cos t_1 )^2 + 4x^2 \sin^2 (t_1) }{ 4x^2 \sin^2 (t_1) } \cdot \frac{x}{2} \cdot(-2) = -x \cdot  \frac{ 4x^2 + 2 \cos t_1 (1-2x^2)   }{ 4x^2 \sin^2 (t_1) } \\
%%%
\Rightarrow &  
\left[ 4x^2(x^2-1) \right]  \cos^2(t_1) +
\left[ -2 (2x^2-1)^2 \right] \cos(t_1) + 
(2x^2-1)^2 = 0
\end{aligned}
\label{eqn:optimal_t1}
\eeq 
Eq.~\eqref{eqn:optimal_t1} gives the optimal $t_1$. Solving Eq.~\eqref{eqn:optimal_t1} gives $\cos (t_1) = \frac{-1+2x^2}{2x^2}$  or $\frac{-1+2x^2}{2(-1+x^2)}$. We take the positive solution which corresponds to a smaller $t_1$. The time-optimal solution $(t^*_1, t^*_2)$ is therefore
\beq 
\begin{aligned}
&\cos (t^*_1 ) = \frac{-1+2x^2}{2(-1+x^2)}, \,\,\,
\tan( \frac{x\, t^*_2 }{2}  ) = 
\frac{1-4 x^2 \sin^2 \frac{t^*_1}{2} }{ 2x \sin t^*_1 }.
\end{aligned}
\label{eqn:optimal_t1_direct}
\eeq 
Taking $x=\frac{1}{2}$, we have $\cos(t^*_1) = \frac{1}{3}$ and thus $t^*_1 \approx 0.392 \pi$; $t^*_2 \approx 0.784 \pi$; $t^*_f = 2t^*_1+t^*_2 \approx 1.5673\pi$. For $x=\frac{1}{\sqrt{32}}$,   $t^*_1 \approx 0.339 \pi$ and $t^*_f \approx 5.221\pi$. These values are consistent with our numerical simulations.

%\begin{figure}[ht]
%\begin{center}
%\includegraphics[width=0.4\textwidth]{Uniform_discriete_Grover_N6}
%\includegraphics[width=0.3\textwidth]{optimal_vs_uniform_diff}
%\caption{The trajectories for uniform (blue) and discrete Grover (red) protocols for $x = 1/\sqrt{2^6} = 1/8$. The uniform protocol, which is a singular control, is very close to the optimal bang-singular control. The trajectory of the discrete-Grover protocol can be viewed as bang-bang approximation to the singular control. }
%\label{fig:Bloch_N6}
% \end{center}
%\end{figure}

When $x$ is small, we get $\cos t^*_1  \approx \frac{1}{2} - \frac{x^2}{2} $ from which we identify $t^*_1 = \frac{\pi}{3} + \delta$. Matching the lowest non-vanishing order in $x$, we have 
\beq 
\begin{aligned}
\cos(t^*_1) &= \cos(\frac{\pi}{3} + \delta) 
\approx \frac{1}{2} - \delta \cdot \sin \frac{\pi}{3} =
\frac{1}{2} - \frac{x^2}{2}  \\
\Rightarrow&  \delta = \frac{1}{\sqrt{3}}, \,\,\, t^*_1 = \frac{\pi}{3} + \frac{x^2}{\sqrt{3} }.
\end{aligned}
\eeq 
To get the small-$x$ expansion of $t^*_2$, we use $\sin \theta = \frac{2 \tan(\theta/2) }{1 + \tan^2(\theta/2)}$ to write the second equation in Eq.~\eqref{eqn:optimal_t1_direct} as 
\beq 
\sin (x t^*_2) = 2 \frac{ (1-4 x^2 \sin^2 \frac{t^*_1}{2} ) \cdot (2x \sin t^*_1) }{(1-4 x^2 \sin^2 \frac{t^*_1}{2})^2 + (2x \sin t^*_1)^2} 
\approx 4 x \sin t^*_1 =  2x\sqrt{3}. 
\eeq 
We expand $x t^*_2 \approx \pi + \delta$ to get 
\beq 
\begin{aligned}
\sin (x t^*_2) &= \sin(\pi + \delta ) \approx - \delta =  2x\sqrt{3} \Rightarrow \delta = -2x\sqrt{3} 
\\
\Rightarrow & t^*_2 = \frac{\pi+\delta}{x} = \frac{\pi}{x} - 2 \sqrt{3}.
\end{aligned}
\eeq 
The total time in the small $x$ limit is 
\beq 
2 t^*_1 + t^*_2 = \frac{\pi}{x} + \frac{2 \pi}{3} - 2 \sqrt{3} \approx \frac{\pi}{x}  - 0.436 \pi. 
\label{eqn:dT_asymptotic}
\eeq 
Note $\pi/x$ is the total time needed in the singular protocol \cite{PhysRevA.57.2403}. 
Fig.~\ref{fig:time_optimal}(a)  shows the total computation time for optimal bang-singular-bang control [by evaluating Eqs.~\eqref{eqn:optimal_t1_direct}], the singular control and Grover's algorithm. For $n$ number of qubits, the overlap $x = 2^{-n/2}$. Their asymptotic behaviors are the same; the acceleration is only noticeable when the system is small. The time difference between the singular and the optimal control, given in Fig.~\ref{fig:time_optimal} (b), is indeed consistent with Eq.~\eqref{eqn:dT_asymptotic}. 
%Finally we note that discrete Grover's algorithm can be regarded as a bang-bang approximation to the singular control. This is illustrated in Fig.~\ref{fig:Bloch_N6}.

\subsection{Example of $x=1/2$}

\begin{figure}[ht]
\begin{center}
\includegraphics[width=0.5\textwidth]{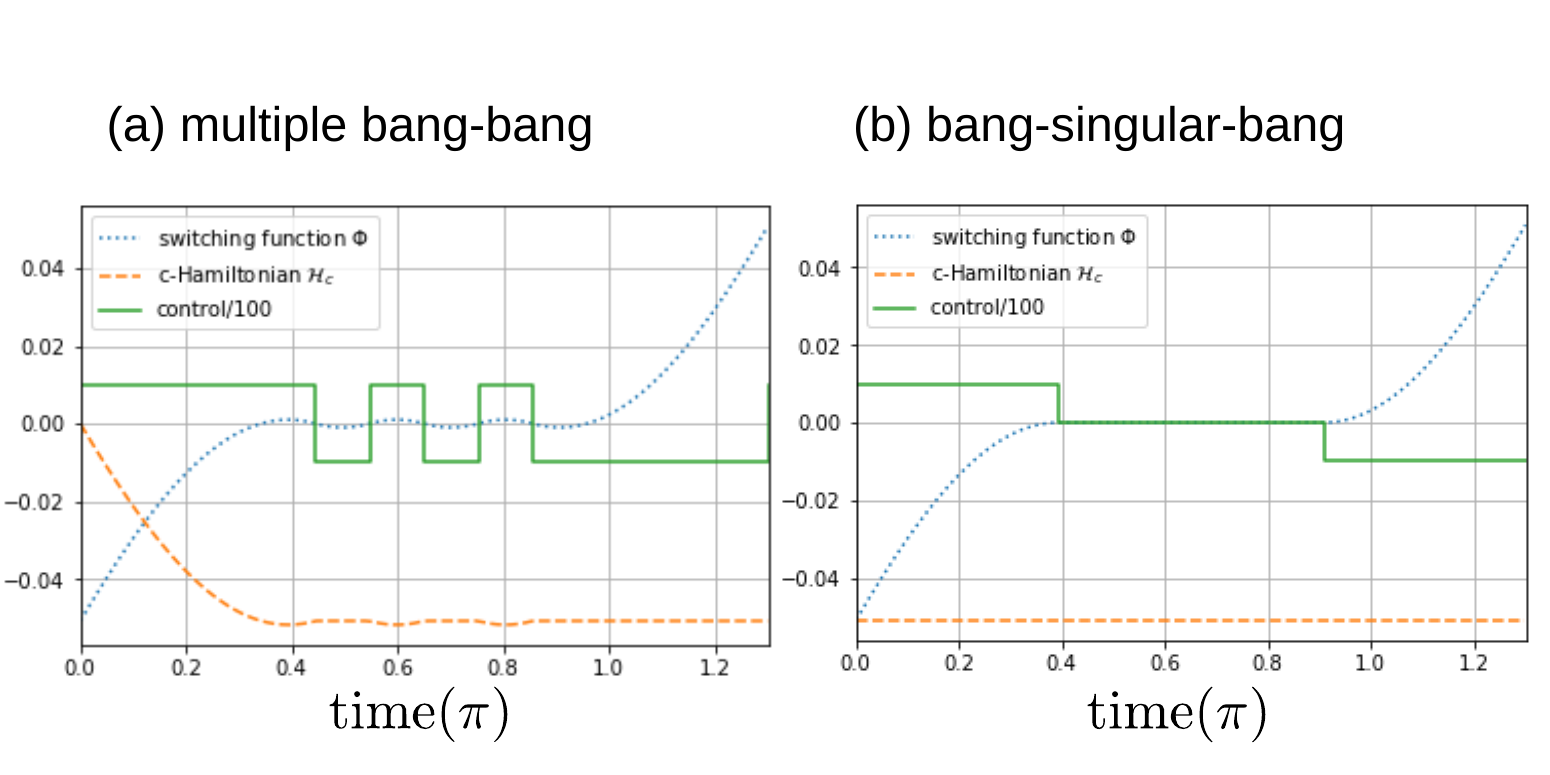}
\caption{ (Color Online) The multiple bang-bang control (a) and the optimal control (b) for $x=\frac{1}{2}$ with the total computation time $t_f = 1.3\pi$, which is not long enough to reach the target state. In (a), the c-Hamiltonian (dashed curve) is not a constant over the time interval, and the switching function (dotted curve) and the control (solid curve) have the same sign, violating the necessary conditions derived from PMP. For the optimal bang-singular-bang control, all necessary conditions are satisfied.
}
\label{fig:Grover_x=1/2_all}
 \end{center}
\end{figure}

More details for the $x=1/2$ case are provided. In particular, we are interested in the behavior for a fixed total time $t_f$ where the total cost function defined in Eq.~\eqref{eqn:J} only has the terminal cost function. The case we will consider is $t_f = 1.3 \pi < t^*_f (x=1/2) \sim 1.57 \pi$, i.e., it is a representative example where the computation time is too short to reach the target state. % We choose $t_f = 1.3 \pi$. 
The optimal bang-singular-bang protocol is  
\beq 
u_{opt}(t; t_1) = \begin{cases}
1 &\text{ for $0<t< t_1 $} \\
0. &\text{ for $t_1 <t< t_f -t_1 $} \\
-1 &\text{ for $t_f -t_1 <t< t_f$}
\end{cases}.
\label{eqn:grover_optimal}
\eeq 
For a given $t_f$, control \eqref{eqn:grover_optimal} only has one parameter $t_1$ which will be determined by numerically minimizing the cost function. We also consider the multiple bang-bang protocol  
\beq 
u_{m}(t; t_1, N) = \begin{cases}
1 &\text{ for $0<t< t_1 $} \\
-1/1 &\text{ alternate $N$ times for $t_1 <t< t_f -t_1 $} \\
-1  &\text{ for $t_f -t_1 <t< t_f$}
\end{cases}.
\label{eqn:grover_int}
\eeq 
Between $t_1 <t< t_f -t_1 $, the control $u$ is alternating between -1 and 1 exactly $N$ times, and each lasts $(t_f - 2 t_1)/(2N)$. For a given $t_f$ and $N$, $t_1$ of control \eqref{eqn:grover_int} will also be determined by numerically minimizing the cost function.

%The trajectories of the control protocols are plotted in Fig.~\ref{fig:Grover_x=1/2_all}. 
Fig.~\ref{fig:Grover_x=1/2_all}(a) shows the switching function $\Phi$ and the c-Hamiltonian $\mathcal{H}_c$ [defined in Eq.~\eqref{eqn:quantum_Hoc_switch}] using the multiple bang-bang control \eqref{eqn:grover_int} with $N=2$ (totally 5 switchings). The numerical optimization gives $t_1 = 0.4446 \pi$. We see that for $t$ between about $0.3 \pi$ and $0.9 \pi$, the control $u(t)$ and the switching function $\Phi(t)$ have the same sign which violates the necessary condition \eqref{eqn:bang_condition}. Also, the c-Hamiltonian $\mathcal{H}_c$  is not a constant over $[0, t_f]$ which violates the necessary condition \eqref{eqn:negative_c-H}.
Fig.~\ref{fig:Grover_x=1/2_all}(b) shows the switching function and the c-Hamiltonian using the optimal bang-singular-bang control \eqref{eqn:grover_optimal}.  The numerical optimization gives $t_1 = 0.3918 \pi$. We see that all necessary conditions are satisfied, i.e., the control $u(t)$ and the switching function $\Phi(t)$ either have opposite signs or are both zero and the c-Hamiltonian $\mathcal{H}_c$ is a constant over the entire $[0, t_f]$. 

\subsection{Summary and discussion on problems of higher dimensions} 
In this section, we have combined the analytical analysis and numerical optimization to show that the control of bang-singular-bang type satisfy all necessary conditions for a time-optimal solution. 
The numerical optimization involves repeating the following three steps: (i) assuming a control $u(t)$; (ii) evolving the wave function from $|\Psi(t=0) \rangle = |s \rangle$ to $|\Psi(t=t_f) \rangle$ (by integrating a differential equation,  a two-dimensional Schr\"odinger equation here); (iii) evaluating the cost function. To check the necessary conditions, we need to (iv) compute the conjugate momentum $|\Pi(t) \rangle$ backwards in time from $t_f$. If $u(t)$ contains many parameters, satisfying the necessary conditions during the entire computation time is practically impossible. Knowing the structure of the control is of great numerical value -- it not only significantly decreases the optimization time by reducing the dimension of search domain but also finds a better solution (i.e., smaller cost function). In Section \ref{sec:geometric_control} we show that the structure of optimal control can be determined even without integration for this problem. 

For problems of higher dimensions, the structure of optimal control by itself can easily become prohibitively difficult to determine. Although the analytical analysis may not be possible, PMP can still be a valuable numerical tool. Specifically, by solving the  Schr\"odinger equation two times for a given $u(t)$ -- one for the wave function $|\Psi(t) \rangle$ [step (ii)] and one for the conjugate momentum $|\Pi(t) \rangle$ [step (iv)], we are able to compute the switching function that reflects the gradient of the cost function with respect to the given $u(t)$. This is known as the ``adjoint state method'' \cite{book:Luenberger}. The gradient obtained using adjoint state method is exact (up to a numerical error), time efficient, and can be used in an iterative optimization algorithm to update $u(t)$.

%%%%%%%%%%%%%%%%%%%%%%%
%%%%%%%%%%%%%%%%%%%%%%%%
\section{Geometric control \label{sec:geometric_control}}

The Schr\"odinger equation in Eq.~\eqref{eqn:dynamics_grover_explicit} involves four (dependent) real variables, and directly applying PMP to a dynamical system of four variables is not immediately informative. 
In this section, we reduce the number of variables to two, and apply geometric control to derive the structure of the time-optimal solution. Grover's algorithm will also be analyzed within this framework.

\subsection{Dimension reduction and Bloch sphere representation}
In view of control theory, a two-dimensional wave function has four (real-valued) dynamical variables. However, because the norm of the wave function is conserved and a global phase is not observable, only two dynamical variables are relevant. These two variables can be parametrized using two angles $(\theta, \phi)$ by expressing a general qubit state as
\beq 
| \Psi(\theta, \phi, \phi_0) \rangle 
= e^{i \phi_0 } 
\begin{bmatrix} \cos \frac{\theta}{2} \\ \sin \frac{\theta}{2} e^{i \phi}  \end{bmatrix} 
\rightarrow 
| \Psi(\theta, \phi) \rangle = \begin{bmatrix} 
\cos \frac{\theta}{2} \\ \sin \frac{\theta}{2} e^{i \phi}  \end{bmatrix} 
\eeq 
with $\theta \in [0, \pi]$ and $\phi \in [0, 2\pi]$. Neglecting the global phase $\phi_0$, a given qubit state corresponds to a point on a unit sphere $S^2$: 
\beq  
( \sin \theta \cos \phi, \sin \theta \sin \phi, \cos \theta )
\Leftrightarrow 
| \Psi(\theta, \phi) \rangle = \begin{bmatrix} 
\cos \frac{\theta}{2} \\ \sin \frac{\theta}{2} e^{i \phi}  \end{bmatrix} 
\eeq
This is the ``Bloch sphere'' representation. In this convention, the first component $\cos \frac{\theta}{2}$ is always real and positive.

%%%%
To investigate the dynamics in $(\theta, \phi)$ manifold, we first need to translate the Hamiltonians in the Schr\"odinger equation to the corresponding vector fields. Since any $2 \times 2 $ Hermitian matrix is a linear combination of three Pauli matrices and the identity matrix, we need to determine the vector fields corresponding to these matrices:
\beq 
\begin{aligned}
\sigma_z &\rightarrow V_z = 2 \partial_\phi
\\ %%%
\sigma_x &\rightarrow V_x = -2 \sin\phi \, \partial_\theta 
- 2 \cos \phi \cot\theta \,  \partial_\phi 
\\ %%
\sigma_y &\rightarrow V_y = 2 \cos\phi \, \partial_\theta  
-2 \sin \phi  \cot \theta \, \partial_\phi.
\end{aligned}
\label{eqn:vector_pauli}
\eeq 
$\{\partial_\theta, \partial_\phi\}$ form a complete basis on the tangent space of $(\theta, \phi)$ manifold. Note that the following commutation relations hold:
\beq 
[V_z, V_x] = -2 V_y, \,\, [V_y, V_z] = -2 V_x, 
\,\, [V_x, V_y] = -2 V_z, 
\eeq 
and the identity matrix that generates a global phase to the qubit wave function in the Schr\"odinger equation has no action on the dynamical variables $(\theta, \phi)$.
The details of Eqs.~\eqref{eqn:vector_pauli} are provided in the Appendix. Using Eqs.~\eqref{eqn:vector_pauli}, the vector fields in Eqs.~\eqref{eqn:H_grover} are
\begin{subequations} 
\begin{align}
H_Y = H_w & \rightarrow \mathbf{Y} = \frac{1}{2} V_z = \partial_\phi 
\label{eqn:vectorX} \\
H_X = H_s %= \frac{1}{2} \left[ (2x \sqrt{1-x^2} ) \sigma_x  + (2x^2-1) \sigma_z \right]  \nonumber \\
&\rightarrow \mathbf{X} 
= -2  x\sqrt{1-x^2} \sin\phi \, \partial_\theta + 
\left[ (2x^2-1) -2 x\sqrt{1-x^2} \cos \phi  \cot \theta \right] \partial_\phi 
\label{eqn:vectorY}
\\
%%%%
H_0 = \frac{1}{2}(H_w + H_s) %=  \frac{x}{2} \left[ \sqrt{1-x^2} \sigma_x + x \sigma_z \right] \nonumber \\
& \rightarrow \mathbf{f} = -x \sqrt{1-x^2} \sin\phi \, \partial_\theta + \left[ x^2 - (x \sqrt{1-x^2} ) \cos \phi  \cot \theta \right] \partial_\phi
\\
H_{d} = \frac{1}{2}(H_w - H_s) %= \frac{1}{2} \left[
%-x \sqrt{1-x^2} \sigma_x + (1-x^2) \sigma_z \right] \nonumber \\
& \rightarrow \mathbf{g} = x \sqrt{1-x^2} \sin\phi \, \partial_\theta + \left[ (1-x^2) + x \sqrt{1-x^2}  \cos \phi  \cot\theta \right] \partial_\phi
%%%%%
\end{align}
\end{subequations} 
The control problem formulated in the Schr\"odinger equation [Eq.~\eqref{eqn:dynamics_grover_explicit}] can now be recast in $(\theta, \phi)$:
\beq 
\begin{aligned}
\frac{\dd}{\dd t} \begin{bmatrix} \theta \\ \phi \end{bmatrix} 
&= \begin{bmatrix} -x \sqrt{1-x^2} \sin \phi \\ 
 x^2 - x \sqrt{1-x^2} \cos \phi \, \cot \theta \end{bmatrix}
 + u(t) 
 \begin{bmatrix} x \sqrt{1-x^2} \sin \phi \\ 
(1-x^2) + x \sqrt{1-x^2} \cos \phi \, \cot \theta \end{bmatrix} \\
%%%%%
&\equiv \mathbf{f}(\theta, \phi) + u \cdot \mathbf{g}(\theta, \phi).
\end{aligned}
\label{eqn:dynamics_theta_phi}
\eeq 
The initial and target values of dynamical variables are 
\beq 
\begin{aligned}
(\theta_i, \phi_i) &= (2 \arctan \frac{\sqrt{1-x^2} }{x}, 0),  \\
%\label{eqn:initial_theta} \\
\theta_f =0 , & \phi_f \text{ any value}.
%\label{eqn:final_theta} 
\end{aligned}
\label{eqn:theta_phi_boundary}
\eeq 
Although Eq.~\eqref{eqn:dynamics_theta_phi} becomes highly non-linear [compared to Eq.~\eqref{eqn:dynamics_grover_explicit}], it is an equation involving only two variables. A two-dimensional control-affine problem allows us to essentially enumerate all possibilities and thus determine the structure of optimal control.

\subsection{General structure of time-optimal solution for two-dimensional systems \label{Sec:singular}}

\begin{figure}[ht]
\begin{center}
\includegraphics[width=0.5\textwidth]{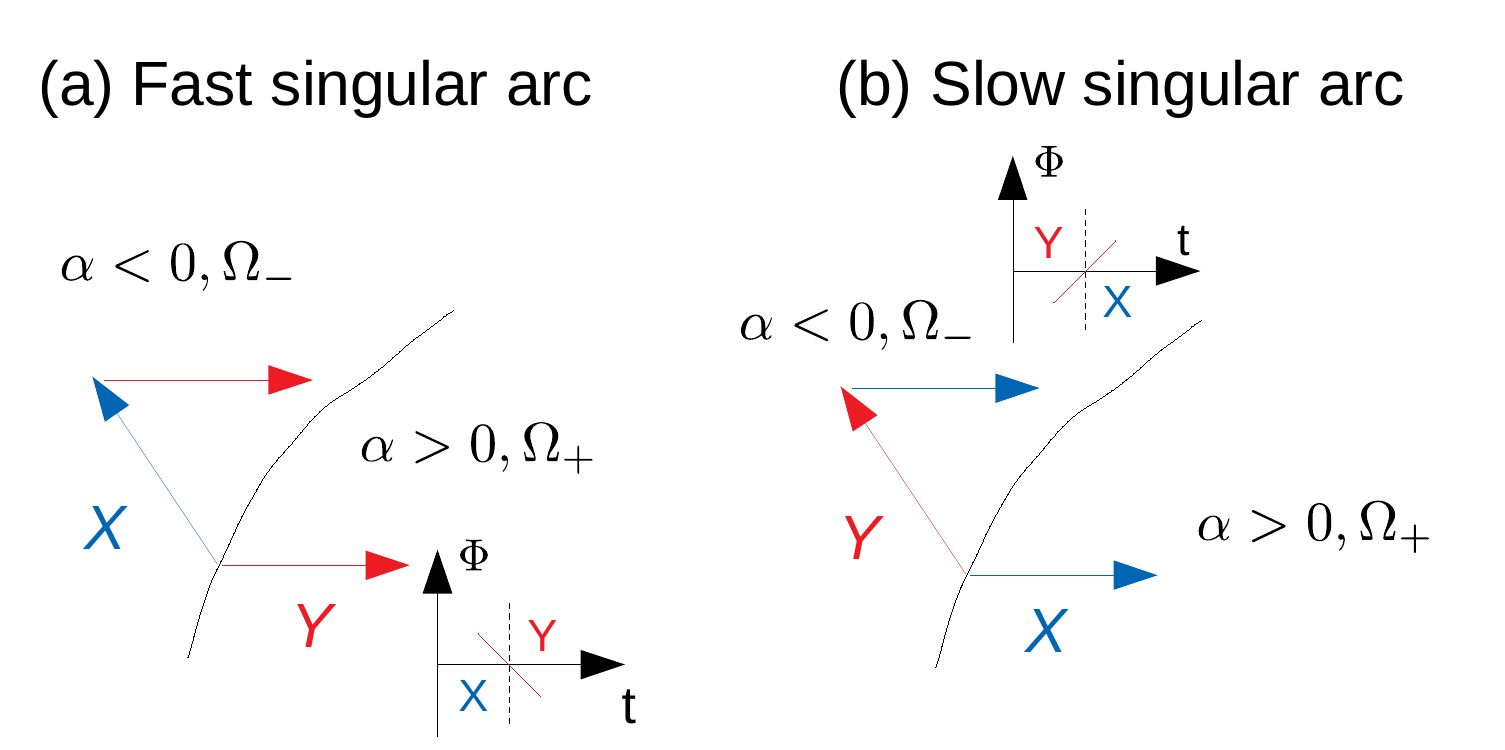}
\caption{(Color Online)  The singular arc $\alpha=0$ divides the manifold into $\alpha>0$, $\Omega_+$ and $\alpha<0$, $\Omega_-$ region. (a) The fast singular arc  where the singular arc represents the optimal-time trajectory. (b) The slow singular arc where the singular arc is not the optimal-time trajectory.  $X$ and $Y$ represent the $u=-1$ and $u=+1$ bang controls respectively. The insets show the behavior of $\Phi(t) \sim 0$, based on which the allowed bang-bang controls in $\Omega_+$ ($YX$) and $\Omega_-$ ($XY$) regions are indicated. The trajectory can be kept around $\alpha=0$ by alternating the $X$ and $Y$ controls.
}
\label{fig:fast_slow_arc}
 \end{center}
\end{figure}

In this subsection, we describe the general structure of the time-optimal solution in two dimension given by Sussmann\cite{Sussmann_87_01, Sussmann_87_02, Sussmann_87_03}. Let us consider the question: if $u(t)$ is an optimal solution, i.e., {\em all} conditions in \eqref{eqn:necc_cond_state_variables} are satisfied but with a zero switching function at time $t$ (i.e., $\Phi(\pmb{x}(t))=0$), what are the possible values of the dynamical variables at a later time $t+\dd t$? The answer relies on $\dot{\Phi}(t)$. Assume that $\mathbf{f}$ and $\mathbf{g}$ are linearly independent, their commutator can be expressed as a linear combination of $\mathbf{f}$ and $\mathbf{g}$, i.e., 
\beq 
[\mathbf{f},\mathbf{g}] = \alpha(\pmb{x})  \mathbf{f} + \beta(\pmb{x}) \mathbf{g},
\eeq 
where $\alpha(\pmb{x})$, $\beta(\pmb{x})$ are smooth functions of $\pmb{x}$. Time derivative of $\Phi(t)$, $\dot{\Phi}(t)$, can be computed as 
\beq 
\begin{aligned}
\dot{\Phi} &= \langle \pmb{\lambda}, [\mathbf{f},\mathbf{g}] (\pmb{x}) \rangle \\
&= \alpha \langle \pmb{\lambda}, \mathbf{f} \rangle + 
\beta \Phi, 
\end{aligned}
\label{eqn:switch_dot}
\eeq 
where $\Phi = \langle \pmb{\lambda}, \mathbf{g} \rangle$ is used. Because $\Phi(t) = 0$ and condition \eqref{eqn:negative_c-H} indicates $\langle \pmb{\lambda}, \mathbf{f} \rangle = -|C|$, we get 
\beq 
\dot{\Phi} = -\alpha \cdot |C|. 
\label{eqn:Phi_dot}
\eeq 
Using Eq.~\eqref{eqn:bang_condition}, we conclude 
\beq 
\begin{cases}
\alpha>0 \Rightarrow \dot{\Phi} <0 & \text{allow } u=-1 \text{ to } u=+1,  XY \text{ control}
\\
\alpha<0 \Rightarrow \dot{\Phi} >0 & \text{allow } u=+1 \text{ to } u=-1,  YX \text{ control}
\end{cases}
\label{eqn:alpha_non-zero}
\eeq 
Here $u=-1$ is denoted as $X$ control; $u=+1$ as $Y$ control. %It is worth noting that in a two-dimensional manifold, $\alpha=0$ represents a curve that divides the neighboring region into $\alpha>0$ ($\Omega_+$) and $\alpha<0$ ($\Omega_-$) domains. 
When $\alpha \neq 0$, $\Phi(t)=0$ can only be an isolated point in time and therefore the optimal control has to be of the bang-bang type. Eq.~\eqref{eqn:alpha_non-zero} is illustrated in Fig.~\ref{fig:fast_slow_arc}. 
 
The next step is to investigate if the admissible control can keep the dynamical variables along $\alpha=0$ that divides the manifold into $\alpha>0$, $\Omega_+$ and $\alpha<0$, $\Omega_-$ regions. If this happens, then $\alpha=0$ defines a singular arc along which the optimal control can be singular ($\Phi(t)=0$ over a finite amount of time); otherwise, the optimal control is of the bang-bang type. To answer this question, we consider $\alpha(\pmb{x}_0) = 0$ and compute the Lie derivative of $\alpha$ with respect to the vector fields $\mathbf{X}$ and $\mathbf{Y}$. Two scenarios arise: 
\bi
\item (i) If $L_\mathbf{X} \alpha(\pmb{x}_0)$ and $L_\mathbf{Y} \alpha(\pmb{x}_0)$ have the same sign, then $\pmb{x}_0$ will move out of $\alpha=0$ for all admissible controls, so $\alpha=0$ is not a singular arc. 

\item (ii) If $L_\mathbf{X} \alpha(\pmb{x}_0)$ and $L_\mathbf{Y} \alpha(\pmb{x}_0)$ have opposite signs, then $\alpha=0$ represents a singular arc. In this scenario, one can keep the dynamical variables around $\alpha=0$ by alternating the $X$ and $Y$ controls.
\ei 

The scenario (ii) has two possible cases. Case (a): if $L_\mathbf{X} \alpha(\pmb{x}_0)<0, L_\mathbf{Y} \alpha(\pmb{x}_0)>0$, the $\alpha=0$  is the time-optimal trajectory; $\alpha=0$ is referred to as a ``fast'' singular arc [see Fig.~\ref{fig:fast_slow_arc}(a)]. In this case, the optimal control is obtained by $d\alpha = 0 = L_\mathbf{f}\alpha + u L_\mathbf{g} \alpha$:
\beq 
u = \frac{ L_\mathbf{X} \alpha(\pmb{x}_0)+ L_\mathbf{Y} \alpha(\pmb{x}_0)  }{  L_\mathbf{X} \alpha(\pmb{x}_0)- L_\mathbf{Y} \alpha(\pmb{x}_0) } 
\label{eqn:optimal_control}
\eeq 
Case (b): if $L_\mathbf{X} \alpha(\pmb{x}_0)>0, L_\mathbf{Y} \alpha(\pmb{x}_0)<0$, the $\alpha=0$  is not the time-optimal trajectory; $\alpha=0$ is referred to as a ``slow'' singular arc [see Fig.~\ref{fig:fast_slow_arc}(b)]. 

The reasoning is provided. In case (a), $\mathbf{X}$ drives $\pmb{x}_0$ to $\Omega_-$ region. To go back to the singular arc, we need to apply $\mathbf{Y}$ but the $XY$ control violates the condition \eqref{eqn:alpha_non-zero}. Therefore, to move along the singular arc, PMP implies that the singular control takes the minimum time. % The same discussion applies if we apply $\mathbf{Y}$ first.
%In $\Omega_-$, only $YX$ control is allowed [Eq.~\eqref{eqn:alpha_non-zero}].
In case (b), $\mathbf{X}$ takes $\pmb{x}_0$ to $\Omega_+$ region. In $\Omega_+$, $XY$ control is allowed by Eq.~\eqref{eqn:alpha_non-zero}, which brings the trajectory back to the singular arc. Therefore, it is not obvious if the bang-bang control or the singular control takes minimum time. It is shown that bang-bang control \cite{book:GeometricOptimalControl} is the optimal solution. It may be a bit surprising that the optimal control can be determined without integrating the differential equation in time (only spatial derivatives needed), and we pinpoint that the condition $\langle \pmb{\lambda}, \mathbf{f} \rangle = -|C|$ leading to Eq.~\eqref{eqn:Phi_dot} is what makes this possible.

\subsection{The structure of optimal solutions to Grover's problem}

\begin{figure}[ht]
\begin{center}
\includegraphics[width=0.8\textwidth]{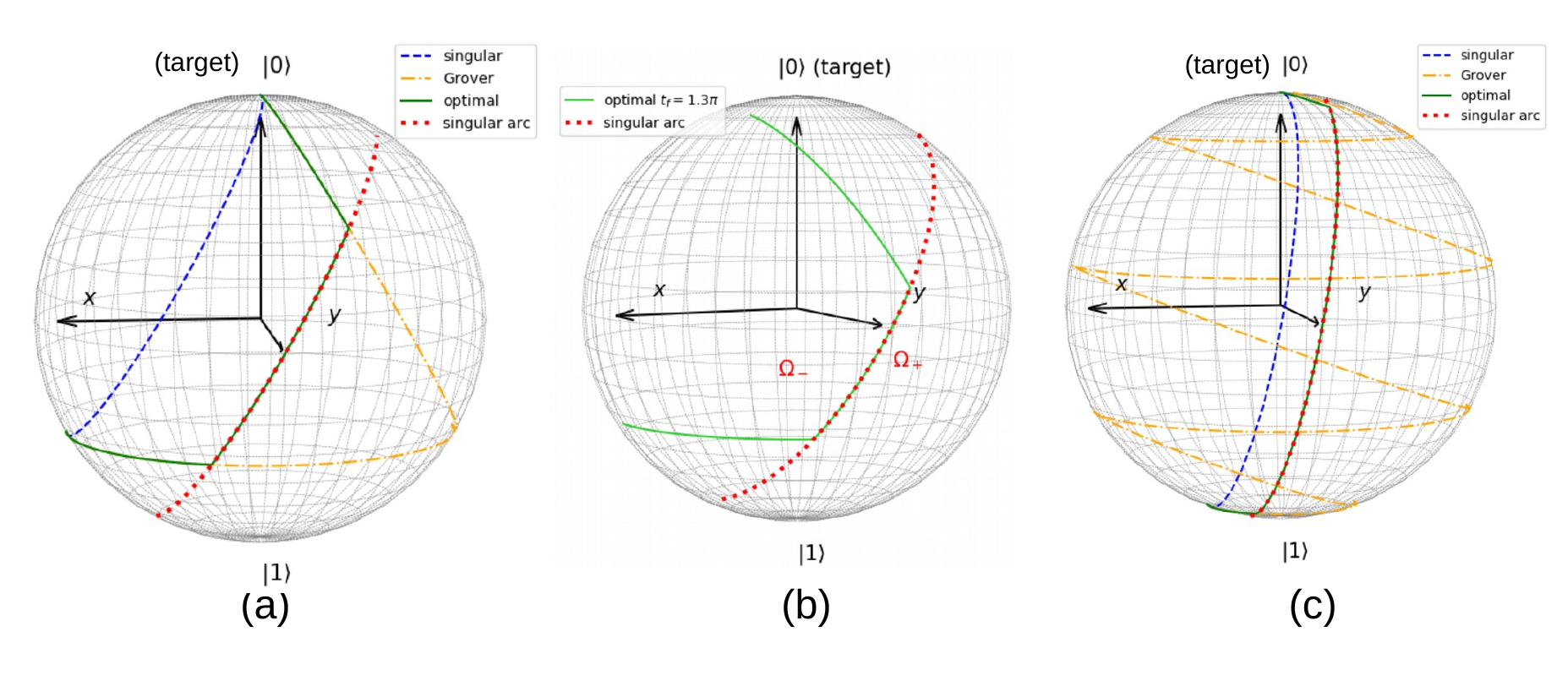}
\caption{(Color Online) Trajectories using the singular (dashed curves), the optimal (solid curves), and Grover's protocol (dash-dot) for (a) $x=\frac{1}{2}$ and $t_f=1.567 \pi$, (b) $x=\frac{1}{2}$ and $t_f=1.3 \pi$ and (c) $x=\frac{1}{4 \sqrt{2}}$ and $t_f= 5.221\pi$. 
The parameters of (a) and (c) are computed from Eq.~\eqref{eqn:optimal_t1_direct}; those of (b) are given in Fig.~\ref{fig:Grover_x=1/2_all}(b). In (b), $t_f = 1.3 \pi$ which is too short to reach the target state. The singular arc (dotted curves) determined by Eq.~\eqref{eqn:singular_arc} is also plotted. The optimal trajectory and the singular arc coincide over a finite amount of time. The trajectory of Grover's algorithm is zigzagging through the optimal trajectory. 
}
\label{fig:singular_arc}
 \end{center}
\end{figure}

The analysis provided in Section \ref{Sec:singular} is now applied to Eq.~\eqref{eqn:dynamics_theta_phi}, where the two-dimensional dynamical variables are $\pmb{x} = (\theta, \phi)$. The commutator $[\mathbf{f},\mathbf{g}]$ is: 
\beq 
\begin{aligned}
[\mathbf{f}, \mathbf{g}] &=  \left[ 
-x \sqrt{1-x^2} \cot \theta \frac{1}{\sin\phi} - (1-x^2)
\frac{\cos \phi}{\sin \phi} \right] \mathbf{f} + \left[ -
x \sqrt{1-x^2} \cot \theta \frac{1}{\sin\phi} + x^2 \frac{\cos \phi}{\sin \phi} \right] \mathbf{g} \\
%%%
&\equiv \alpha(\theta, \phi) \mathbf{f} + \beta (\theta, \phi) \mathbf{g}.
\end{aligned}
\label{eqn:[f,g]}
\eeq 
The curve defined by $\alpha(\theta, \phi) =- \sqrt{1-x^2} (x \frac{\cot \theta}{\sin \phi} + \sqrt{1-x^2} \frac{\cos \phi}{ \sin \phi} ) = 0$ is 
\beq 
\begin{aligned}
%&\alpha(\theta, \phi) = 0 %-\frac{\sqrt{1-x^2}}{ \sin \phi} \left[ x  \cot \theta + \sqrt{1-x^2} \cos \phi \right] =0 \\
%\Rightarrow  
-\frac{x}{ \sqrt{1-x^2} } \cot \theta = \cos \phi. 
\end{aligned}
\label{eqn:singular_arc}
\eeq 
This curve divides the region around $\alpha=0$ into $\alpha>0$ ($\Omega_+$) and $\alpha<0$ ($\Omega_-$) regions. To identify these regions, we note that $\alpha(\frac{\pi}{2}, \frac{\pi}{2})=0$, and $\alpha(\frac{\pi}{2}, \frac{\pi}{2}+ \epsilon)>0$, $\alpha(\frac{\pi}{2}, \frac{\pi}{2} - \epsilon)<0$ (with $\epsilon$ a small positive number).  

%The tangent vector of this curve is determined by 
%\beq 
%\begin{aligned}
%& -\frac{x}{ \sqrt{1-x^2} } \frac{ - d\theta}{ \sin^2 \theta} = -\sin  \phi \, d\phi \\ \Rightarrow   & \frac{d \theta}{d \phi} = - \frac{ \sqrt{1-x^2} }{ x} \sin \phi \sin^2 \theta\,\, \left(\text{use } \sin^2 \theta = \frac{ x^2 }{ x^2 + (1-x^2) \cos^2 \phi } \right) \\
%& = \frac{ -x \sqrt{1-x^2} \sin \phi }{ x^2 + (1-x^2) \cos^2 \phi }
%= \frac{ V_{u=0.5, \theta} }{ V_{u=0.5, \phi} }. 
%\end{aligned}
%\eeq 

To see if $\alpha=0$ corresponds to a singular control, we compute the Lie derivative of $\alpha(\theta, \phi)$ with respect to $\mathbf{Y}$ and $\mathbf{X}$. Using $\partial_\theta \alpha = \frac{ x \sqrt{1-x^2} }{ \sin \phi\sin^2 \theta}$ and $\cot \theta = - \frac{\sqrt{1-x^2}}{x} \cos \phi$ (the condition of $\alpha=0$), we get
\beq 
\begin{aligned}
 L_{\mathbf{Y}} \alpha &=  \partial_\phi \alpha 
%\frac{ \sqrt{1-x^2} }{ \sin^2 \phi} \left[ x \cot \theta \cos \phi + \sqrt{1-x^2} \right] = 
%\frac{1-x^2}{ \sin^2 \phi} \left[ 1- \cos^2 \phi \right]
= 1-x^2 >0, \\
L_{\mathbf{X}} \alpha &= %\left[ -2  x\sqrt{1-x^2} \sin\phi \, \partial_\theta +  \left[ (2x^2-1) -2 x\sqrt{1-x^2} \cos \phi  \cot \theta \right] \partial_\phi \right] \alpha \\
%& = -2  x\sqrt{1-x^2} \sin\phi \, \partial_\theta \alpha + 
%\left[ 2x^2 \sin^2 \phi + 2 \cos^2 \phi -1 \right] \partial_\phi \alpha 
-(1-x^2) = -L_{\mathbf{Y}} \alpha < 0. 
\end{aligned}
\label{eqn:Lie_Grover}
\eeq 
Substituting Eq.~\eqref{eqn:Lie_Grover} into Eq.~\eqref{eqn:optimal_control}, we see that $\alpha = 0$ corresponds to a singular control with $u=0$. 
Fig.~\ref{fig:singular_arc} gives the trajectories using the singular, the optimal, and Grover's protocol for $x=1/\sqrt{4}$ and $x=1/\sqrt{32}$. Note that in Fig.~\ref{fig:singular_arc}(b) we purposely use $t_f = 1.3 \pi$ which is too short to reach the target state [see Fig.~\ref{fig:Grover_x=1/2_all}(b)]. The singular arc determined by Eq.~\eqref{eqn:singular_arc} is also plotted. The optimal trajectory does coincide with the singular arc over a finite interval of time. 

%%%%%%%%%%%%%%%%%%%%%%%%%%%%
\subsection{Relation to the quantum state preparation}

Although we only focus on Grover's problem, the same technique is applicable to problems which can be reduced to a single qubit and some qualitative behaviors may provide insights into problems of higher dimensions. In this subsection we discuss two concrete examples related to the quantum state preparation. 

For the Grover's problem, the structure of singular arc and the position of the initial state (see Fig.~\ref{fig:singular_arc}) naturally introduce two critical times $T_1$ and $T_2 (>T_1)$. For the computation time  $t_f<T_1$, the trajectory does not touch the singular arc and the optimal control is of bang-bang type ($Y$-$X$). $T_2$ is defined as the minimum time where the target state can be reached. For $T_1< t_f< T_2$, the optimal control is of bang-singular-bang type ($Y$-0-$X$). For $t_f>T_2$, there is no unique trajectory to reach the target state. If only the bang-bang control is allowed, there will be many switchings along the singular arc. These behaviors are found in the quantum state preparation problem (for both single-qubit and multi-qubit systems) studied in Refs.~\cite{PhysRevX.8.031086, PhysRevLett.122.020601}. The fact that a singular control can be approximated by many bang-bang controls is reflected as the ``glassy'' phase (in time domain) in Refs.~\cite{PhysRevX.8.031086, PhysRevLett.122.020601}. 

For the second example, we consider the time-optimal control that steers a two-qubit product state to the fully entangled state, which is formulated and solved in Ref.~\cite{PhysRevA.97.062343}. The dimension of the two-qubit Hilbert space is four. Using the symmetry of the Hamiltonian, the four-dimensional Hilbert space can be divided into two invariant subspaces and each subspace is effectively a single qubit. The formalism provided in Section II and IV.A can then be applied to one of the subspace. Without providing details, our analysis does show that the singular-bang control is the time-optimal solution as found in Ref.~\cite{PhysRevA.97.062343}. One benefit of geometric control is that we are able to compute the value of the singular control [using Eq.~\eqref{eqn:optimal_control}] without performing time integration.

\subsection{Analysis of Grover's algorithm in the reduced dimension}
We conclude this section by analyzing Grover's algorithm in the  $(\theta, \phi)$ manifold, where the problem is to steer the dynamical variables from $(\theta_i, \phi_i) = (2\arctan \frac{\sqrt{1-x^2} }{x}, 0)$ to $\theta_f = 0$. We first note that by substituting $\tan \theta_i = \frac{\sqrt{1-x^2} }{x}$ and $\phi_i=0$ into Eq.~\eqref{eqn:vectorX}, one gets $\mathbf{X}(\theta_i, \phi_i) = 0$. 
This indicates that the $X$ ($u=-1$) control cannot change the initial dynamical variables  $(\theta_i, \phi_i)$ and the bang control has to start with $Y$ ($u=+1$). Next, we notice that %the ultimate goal is to reduce $\theta$ from $\theta_i$ to 0 and 
only one of the bang controls, the $X$ control, can change the value of $\theta$.
% $Y$ control can only alter $\phi$. 
To drive $\theta$ from $\theta_i$ to 0 with a minimum number of bang-to-bang switchings, we need to maximize the reduction of $\theta$ during the $X$ control. To obtain the maximum $\theta$-reduction, we consider the dynamics with the $X$ control given in Eq.~\eqref{eqn:vectorX}:
\begin{subequations} 
\begin{align}
 \frac{\dd \theta}{\dd t} &= -2x \sqrt{1-x^2} \sin \phi, 
 \label{eqn:Grover_theta}\\
 \frac{\dd \phi}{\dd t} &= (2x^2-1) -2 x\sqrt{1-x^2} \cos \phi  \cot \theta.
 \label{eqn:Grover_phi}
\end{align}
\end{subequations} 
For small $x$, $x \cot \theta \leq 1/N$ holds for $\theta \in [Nx, \pi-Nx] $ (one can pick $N =10$), indicating $\frac{\dd \phi}{\dd t} \approx -1$  [Eq.~\eqref{eqn:Grover_phi}] over this range. %Fig.~\ref{fig:singular_arc}, we see that the optimal trajectory somehow keeps $\phi$ around $\pi/2$ which maximizes the decreasing rate of $\theta$. 
Integrating Eq.~\eqref{eqn:Grover_theta} from $\phi_i$ to $\phi_f$ with $\frac{\dd \phi}{\dd t} \approx -1$, we get the change of $\theta$ as 
\beq 
\begin{aligned}
\Delta \theta &= -2x \sqrt{1-x^2} \int_{\phi_i}^{\phi_f} 
\sin \phi \frac{\dd t}{\dd \phi} \dd \phi \\
%&\approx +2x \sqrt{1-x^2} \int_{\phi_i}^{\phi_f} \sin \phi \,d\phi \\
&\approx 2x \sqrt{1-x^2} \left[ \cos \phi_i - \cos \phi_f \right] \\
&\geq 2x \sqrt{1-x^2} \left[ - 2  \right]
\equiv -|\Delta \theta|_\text{max}
\end{aligned}
\eeq 
The maximum $\theta$-reduction $|\Delta \theta|_\text{max} = 4x \sqrt{1-x^2}$ can only be obtained when $\phi_i = \pi$ and $\phi_f = 0$. To achieve $|\Delta \theta|_\text{max}$ from $\phi_i=0$, one first (a) applies $Y$ ($\mathbf{Y} = \partial_\phi$) control for time $\pi$ that takes $\phi$ from 0 to $\pi$, and then (b) applies $X$ ($\mathbf{X} \approx -2x \sin\phi \partial_\theta-\partial_\phi$ for small $x$) control for time $\pi$ that takes $\phi$ from $\pi$ back to 0; it is (b) that reduces the $\theta$ by $|\Delta \theta|_\text{max}$. (a) and (b) will be repeated $N$ times to bring $\theta=\theta_i$ to $\theta=0$. For small $x$, $\theta_i \approx \pi$ and $N$ is estimated as
\beq 
\frac{\theta_i}{|\Delta \theta|_\text{max} } \approx \frac{\pi}{4x}.
\eeq
This procedure is identical to Grover's algorithm described in Section \ref{sec:existing}. We conclude that in terms of optimal control, Grover's algorithm uses the bang-bang protocol with a minimum number of switchings to approximate the optimal bang-singular-bang control. To visualize this point, in Fig.~\ref{fig:singular_arc}(c) we show that the trajectory of Grover's algorithm is zigzagging through the optimal trajectory. Cutting the number of switchings reduces the query complexity, which is how the quantum speedup is defined in quantum computation using a discrete sequence of gates.

%%%%%%%%%%%%%%%%%%%%%%%%%
\section{Conclusion}

Grover's quantum search problem is one of the most well known and important problems in quantum computation. 
In this paper, we formulate this problem as a time-optimal control problem, and apply Pontryagin's Minimum Principle to solve it. 
Designing a quantum algorithm is equivalent to solving a time-optimal control problem where the dynamical variables are the components of the wave function, the dynamics are the specified by the Schr\"odinger's equation, and the target state is the ground state of the given ``problem Hamiltonian''. 
%%%%%
Although the dimension of the Hilbert space of Grover's problem can be arbitrarily large, it can be reduced to a two-dimensional Schr\"odinger equation.
We show that the optimal control for Grover's problem has the bang-singular-bang structure % by checking all necessary conditions imposed by Pontryagin's Minimum Principle.  
and %analytically determine the minimum time and
explicitly demonstrate that the optimal protocol does take shorter time to reach the target state when comparing to other known protocols.
This bang-singular-bang can be derived  without integrating the differential equations by using geometric control technique. The key step is to derive the dynamical equations involving only two real-valued dynamical variables.  Within this manifold, applying geometric control provides the optimal protocol at each point. The formalism provided in this paper can be beneficial to any quantum systems which can be  effectively described by a single qubit. We are aware that the generalization to systems of higher dimensions may not be as informative as the analysis relies on enumerating all possibilities, but can be worth investigating. The key insight brought by time-optimal control theory is that in addition to the bang-bang protocol the singular control should be considered. 
However, PMP provides an efficient and exact formalism to compute the gradient of the cost function at a given control, which can be useful in numerical optimization for problems of higher dimensions. 
In view of optimal control, Grover's algorithm uses bang-bang protocol, with a minimum number of bang-to-bang switchings to reduce the query complexity, to approximate the optimal protocol. Our work provides a concrete example how optimal control is connected to the quantum computation, and may shed some light on how a  quantum algorithm can be designed. 

\section*{Acknowledgment}
C.L. thanks Yanting Ma, Helena Zhang and Andrew Millis for very helpful discussions. We thank  Dries Sels for very valuable comments.  %We thank ......

%%%%%%%%%%%%%%%%%%%%%%%%%
\appendix 

\section{Operator as a vector field in $(\theta, \phi)$}
In this appendix, Eq.~\eqref{eqn:vector_pauli} that relates the vector fields defined on the tangent space of $(\theta, \phi)$ to the Pauli matrices are derived.  
Applying $H = \sigma_z$ on a state  $| \Psi(\theta, \phi) \rangle$ over a time interval $d t$, we get 
\beq 
e^{-i \sigma_z \cdot \dd t}
\begin{bmatrix} 
\cos \frac{\theta}{2} \\ \sin \frac{\theta}{2} e^{i \phi}  \end{bmatrix} 
= \begin{bmatrix} 
\cos \frac{\theta}{2} e^{-i \,\dd t} \\ 
\sin \frac{\theta}{2} e^{i (\phi+\dd t)}  \end{bmatrix}
\rightarrow 
\begin{bmatrix} 
\cos \frac{\theta}{2} \\ \sin \frac{\theta}{2} e^{i (\phi + 2\,\dd t)}  \end{bmatrix}.
\eeq 
After forcing the first component to be real and positive, $e^{-i \sigma_z \cdot \dd t}$ only changes the $\phi$ component. 
We therefore conclude $\sigma_z \rightarrow 2 \partial_\phi$. 

%%%
Applying $H = \sigma_x$ on a state  $| \Psi(\theta(0), \phi(0)) \rangle$ over a time interval $d t$, we get
\beq 
\begin{aligned}
& e^{-i \sigma_x \cdot \dd t}
\begin{bmatrix} 
\cos \frac{\theta(0)}{2} \\ \sin \frac{\theta(0)}{2} e^{i \phi(0)}  \end{bmatrix} 
= \begin{bmatrix} 1 & -i\,\dd t \\ -i \,\dd t & 1 \end{bmatrix} 
\begin{bmatrix} 
\cos \frac{\theta(0)}{2} \\ \sin \frac{\theta(0)}{2} e^{i \phi(0)}  \end{bmatrix} \\
%%%
=& \begin{bmatrix} 
\cos \frac{\theta(0)}{2}  + \dd t \, \sin \frac{\theta(0)}{2} \sin \phi(0)
-i \,\dd t\, \sin \frac{\theta(0)}{2} \cos \phi(0) \\ 
\sin \frac{\theta(0)}{2} e^{i \phi(0)} - i \,\dd t \cos \frac{\theta(0)}{2}
\end{bmatrix}
\rightarrow 
\begin{bmatrix} 
\cos \frac{\theta(\dd t) }{2} \\ \sin \frac{\theta(\dd t)}{2} e^{i \phi(\dd t)}  \end{bmatrix}.
\end{aligned}
\eeq 
Applying $\sigma_x$ changes both $\theta$ and $\phi$ components. We note that $A + i \,\dd t = A (1 + i (\dd t/A) ) \sim A e^{i \dd t/A}$, so to the first order a small imaginary part does not change the amplitude. For the first component, we see that 
\beq 
\begin{aligned}
&\cos \frac{\theta(\dd t) }{2} = \cos \frac{\theta(0)}{2}  + \dd t \, \sin \frac{\theta(0)}{2} \sin \phi(0) \\ 
%%%
%\Rightarrow & \cos \frac{\theta(dt) }{2} - \cos \frac{\theta(0)}{2}  
%= -\sin \frac{\theta(0)}{2} \frac{1}{2} \frac{d \theta}{dt} dt
%= dt \, \sin \frac{\theta(0)}{2} \sin \phi(0) 
\Rightarrow & \frac{\dd \theta}{\dd t} = -2 \sin \phi. 
\label{eqn:Vy_theta}
\end{aligned}
\eeq 
The amplitude change of the first component determines the coefficient of $\partial_\theta$. The first component also introduces a phase term $1 - i\,\dd t\, \tan\frac{\theta(0)}{2} \cos \phi(0)$ which has to be compensated. Therefore, the second component gives 
\beq 
\begin{aligned}
& \sin \frac{\theta(\dd t)}{2} e^{i \phi(\dd t)} 
= \left[ \sin \frac{\theta(0)}{2} e^{i \phi(0)} - i \,\dd t \cos \frac{\theta(0)}{2} \right] \left[ 
1 + i\,\dd t\, \tan\frac{\theta(0)}{2} \cos \phi(0)
\right] \\ 
%%%
\Rightarrow & \sin \frac{\theta(\dd t)}{2} e^{i \phi(\dd t)} - \sin \frac{\theta(0)}{2} e^{i \phi(0)} = i\,\dd t \left[ \sin \frac{\theta(0)}{2} \tan \frac{\theta(0)}{2} \cos \phi(0) e^{i \phi(0)} - \cos \frac{\theta(0)}{2}
\right] \\
%%%
\Rightarrow & 
\frac{1}{2} \cos \frac{\theta}{2}  \frac{\dd \theta}{\dd t} 
+ \sin \frac{\theta}{2}  i \frac{\dd \phi}{\dd t} 
= i \left[ \sin \frac{\theta}{2} \tan \frac{\theta}{2} \cos \phi  - \cos \frac{\theta}{2} \, e^{-i \phi} \right]. 
\end{aligned}
\eeq 
Using $\frac{\dd \theta}{\dd t} = -2 \sin \phi$ and $e^{-i \phi} = \cos \phi - i \sin \phi$, we get 
\beq 
\begin{aligned}
&- \cos \frac{\theta}{2} \sin \phi + i \sin \frac{\theta}{2} \frac{\dd \phi}{\dd t} = i \left[ \sin \frac{\theta}{2} \tan \frac{\theta}{2} \cos \phi  - \cos \frac{\theta}{2} \, (\cos \phi - i \sin \phi) \right] \\
%%%
\Rightarrow & \frac{\dd \phi}{\dd t} = \cos \phi \left[ 
\tan \frac{\theta}{2} - \cot \frac{\theta}{2}\right]
= -2 \cos \phi \cot \theta. 
\end{aligned}
\label{eqn:Vy_phi}
\eeq 
Combining Eq.~\eqref{eqn:Vy_theta} and ~\eqref{eqn:Vy_phi}, we get 
\beq 
\sigma_y \rightarrow -2 \sin \phi \partial_\theta - 
2 \cos \phi \cot \theta \partial_\phi.
\eeq 
The vector field corresponding to $\sigma_y$ can be derived similarly. A straightforward calculation gives
\beq 
\begin{aligned}
n_x \sigma_x + n_z \sigma_z &\rightarrow n_x V_x + n_z V_z 
= -2  n_x \sin\phi \, \partial_\theta 
+ 2 \left[  n_z - n_x \cos \phi  \cot \theta \right] \partial_\phi,
\end{aligned}
\label{eqn:vector_field}
\eeq
which is useful to map Hamiltonians appeared in Grover's problem to their corresponding vector fields. 
%%%

%%%%%%%%%%%%%%%%%%%%%%%%%%%%%%%%%%%%%%%%%%%
\bibliography{QC_optimal_control}

\begin{thebibliography}{10}
\bibitem{convention}
Note that the conventional Hamiltonian is given by $H = u(t) H_w + (1-u(t))
  H_s$ with $0<u<1$. To be consistent with the convention in control thoery, we
  define $H_0$ and $H_d$ such that $|u|<1$.
\end{thebibliography}
\bibliographystyle{unsrt}
%\begin{thebibliography}{10}
%\end{thebibliography}

%\appendix

\end{document}